\documentclass[%
preprint,
 amsmath,amssymb,
 aps,
]{revtex4-2}

\usepackage{graphicx}
\usepackage{dcolumn}
\usepackage{bm}
\usepackage{physics}

\begin{document}


\title{Level shifts of exotic deuterium atoms from effective range parameters}
\author{Pierre-Yves Duerinck}
\email{pierre-yves.duerinck@ulb.be}
\affiliation{Physique Nucléaire Théorique et Physique Mathématique, C.P. 229, Université libre de Bruxelles (ULB), B-1050 Brussels, Belgium. \\ IPHC, CNRS/IN2P3, Université de Strasbourg, 67037 Strasbourg, France.}

\author{Rimantas Lazauskas}
\email{rimantas.lazauskas@iphc.cnrs.fr}
\affiliation{IPHC, CNRS/IN2P3, Université de Strasbourg, 67037 Strasbourg, France.}

\author{Jérémy Dohet-Eraly}
\email{jeremy.dohet-eraly@ulb.be}
\affiliation{Physique Nucléaire Théorique et Physique Mathématique, C.P. 229, Université libre de Bruxelles (ULB), B-1050 Brussels, Belgium.}

\date{\today}

\begin{abstract}
We discuss various methods for determining the level shifts induced by short-range interactions in exotic atoms which are formed by a hadron and a nucleus of opposite charge. The energy shifts can be related to the effective range parameters through  Deser or Trueman formulas. In order to assess the accuracy and the reliability of this indirect approach, its predictions are compared with direct bound-state calculations. Notably, the importance of including the Coulomb potential in scattering calculations is highlighted. Finally, the impact of multipole terms in the asymptotic electromagnetic interaction is examined. It is demonstrated that an indirect approach remains relevant by accounting for these effects perturbatively. 

\end{abstract}

\maketitle


\section{Introduction}
Exotic atoms are formed when a low-energy  negatively charged particle is captured by an atom, with the concomitant emission of one of the atomic electrons~\cite{G04}. These systems offer unique opportunities to test fundamental symmetries of the Standard Model as well as properties of strong interaction. In modern experiments, the most common exotic atoms involve particles such as muons, pions, antikaons, or antiprotons.
Upon capture, these particles typically occupy highly excited Coulomb orbitals, but they rapidly decay to lower energy states through a cascade of X-rays or the emission of Auger electrons. Lifetime of these systems is finite either due to negative particle decay, driven by the weak interaction (for muonic systems) or due to the annihilation with a nucleon from the atomic nucleus (for kaonic and antiprotonic systems). The PUMA experiment \cite{PUMA}, for instance, aims to exploit this latter phenomenon for probing the tail of unstable nuclei using low-energy antiprotons.
 
The spectrum of exotic atoms made by a nucleus and a negatively charged particle closely resembles the one of hydrogenic atoms. It includes an infinite set of Rydberg states that are slightly shifted and broadened with respect to hydrogenic levels because the interaction between the exotic particle and the nucleus cannot be reduced to a point-like Coulomb potential. The level shifts of pionic, kaonic, and antiprotonic hydrogen have been measured in various experiments \cite{pionic_hyd,SIDD,G99} and have served as a foundation for modeling interactions between nucleons and these particles. Measurements have also been performed for pionic and antiprotonic deuterium \cite{pionic_deut,G99} while they are planned for kaonic deuterium by the SIDDHARTA collaboration \cite{SIDD_deut}. Exotic deuterium atoms are the simplest systems beyond hydrogen atoms. They provide a valuable testing ground for evaluating the accuracy of nuclear interactions in describing more complex systems.

From the point of view of a microscopic approach, the evaluation of the level shifts via direct bound-state methods is numerically challenging. Due to the presence of two very different physical scales in the system at hand – the size of the nucleus and the size of the Rydberg states – a basis of substantial extension is necessary to describe adequately both short- and long-range interactions. Instead, an indirect calculation of the level shifts may be obtained from effective range parameters using formulas derived by Deser \cite{DGBT54} and Trueman \cite{T61}. This approach has proven to be effective for antiprotonic deuterium \cite{CL21,DLC23}, though it has been limited to $S$ waves and spin-uncoupled $P$ states because of the coupling between different partial waves induced by the deuteron quadrupole moment in coupled $P$ states.
 In the present work, the low-lying states of kaonic and antiprotonic atoms are studied by solving the Faddeev-Merkuriev equations in configuration space. The level shifts are computed from direct and indirect approaches, which allows one to compare the accuracy of Deser and Trueman formulas, which differ from their treatment of the Coulomb interaction. For this purpose, the use of Trueman relation \cite{T61} has been extended to coupled antiproton-deuteron $P$ waves by treating the effects of the quadrupole moment at the first order of perturbation theory. 
 
This paper begins with a review of various formulas that relate level shifts to effective range parameters, with examples provided for a two-body system. The formalism is then extended to three-body calculations, incorporating different particle channels. Finally, we compute the level shifts of kaonic and antiprotonic deuterium using both direct and indirect approaches. In both cases, the dependence of the results on the nucleon-nucleon ($NN$) interaction is investigated.

\clearpage

\section{Level shifts from effective range parameters}
Let us consider the simplest model for an exotic atom: a system of two particles interacting through an attractive Coulomb potential $V_C(r)=-\frac{e^2}{r}$ and a short-range potential $V_{\rm{sr}}(r)$ representing the strong interaction between the nucleus and the negatively charged particle. If the short-range forces are sufficiently attractive, they can generate deep bound states that are only mildly affected by the Coulomb potential. Such states are not considered here. In addition, the spectrum of such system contains an infinite set of Rydberg states, with energies approximately given by the energy of an hydrogen-like atom:
\begin{equation}
\epsilon_n = - \frac{\mu \alpha^2}{2 n^2}, \label{eps_n}
\end{equation}
where $\mu$ is the reduced mass of the system, $\alpha$ is the fine-structure constant, and $n$ is the principal quantum number. Depending on the nature of the short-range forces, these hydrogen-like levels undergo a shift as well as a broadening due to the finite lifetime of the state caused by annihilation, which makes the energy a complex quantity. The level shift is defined as
\begin{equation}
\Delta E_{nl} = E_{nl}-\epsilon_n = \Delta E_R - i \frac{\Gamma}{2},
\end{equation}
where $E_{nl}$ is the energy of the system. The energy of these states can be reliably extracted from the effective range paramters \cite{T61}. A connection between level shift and scattering length  was first established by Deser \textit{et al.} in Ref. \cite{DGBT54}, where the authors showed that the $1s$ level shift can be approximated by
\begin{equation}
\frac{\Delta E^{(D_0)}_{1s}}{\epsilon_1} = - \frac{4 A_0}{B}, \label{D0}
\end{equation}
where $A_0$ is the scattering length defined in the absence of Coulomb field and $B=\frac{\hbar}{\alpha \mu}$ is the Bohr radius. This formula has later been improved in Ref. \cite{DI} to describe kaonic hydrogen. Within the framework of non-relativistic effective field theory, the level shift is expressed in a series expansion, providing the following improved Deser formula:
\begin{equation}
\frac{\Delta E^{(D_I)}_{1s}}{\epsilon_1} = - \frac{4 A_0}{B} \left[1+ \frac{2A_0}{B} \left(\ln \alpha -1 \right) \right]. \label{DIe}
\end{equation}
It turns out that the series leading to Eq. \eqref{DIe} can be resummed \cite{DRS,DRS2}, providing a more accurate expression for the level shift:
\begin{equation}
\frac{\Delta E^{(D_{\Sigma})}_{1s}}{\epsilon_1} = \frac{- 4 A_0}{B- 2 A_0 \left(\ln \alpha -1 \right)}, \label{DRSe}
\end{equation}
referred as resummed Deser formula. When deriving Eqs. \eqref{D0}-\eqref{DRSe}, the interference between the Coulomb and strong interaction was disregarded, which limits the accuracy of the former expressions. A more rigorous treatment of the Coulomb potential was presented by Trueman in Ref. \cite{T61},  based on a modified effective range expansion:
\begin{equation}
k^{2l+1} \left[(2l+1)!! \right]^2 C^2_l(\eta) \cot(\delta_l) + h_l(\eta) = - \frac{1}{a_l} + \frac{1}{2} r_l k^2 + \mathcal{O}(k^4),  \label{effrange}
\end{equation}
where $k$ is the wavenumber, $\delta_l$ is the scattering phase shift, $\eta=-\frac{1}{kB}$ is the Sommerfeld parameter, $a_l$ is the scattering length, and  $r_l$ is the effective range. The coefficients $C_l(\eta)$ and $h_l(\eta)$ are given, for example, in Ref. \cite{CRW92}. Bound or resonant states correspond to poles of the $S$-matrix, and their energy can thus be determined by solving Eq. \eqref{effrange}, neglecting the terms in $k^4$ and beyond, and setting $\cot(\delta_l)=i$. Trueman formally expressed the solution to this problem as a series expansion in $\frac{a_l}{B^{2l+1}}$. Specifically, the level shifts for $S$ and $P$ waves are given, at second order in $\frac{a_l}{B^{2l+1}}$, by \cite{T61,CRW92}
\begin{align}
\frac{\Delta E^{(T)}_{nl}}{\epsilon_n} = - 4 \, \frac{ \alpha_{nl}}{n} \frac{a_l}{B^{2l+1}} \left(1- \beta_{nl} \frac{a_l}{B^{2l+1}} \right),  \label{Tr}
\end{align}
where the coefficients $\alpha_{nl}$ are defined as:
\begin{align}
\alpha_{n0}&=1, \\
\alpha_{nl}&=\prod_{s=1}^l \left(\frac{1}{s^2}-\frac{1}{n^2} \right).
\end{align}
For $l=0$ and $l=1$, the coefficients $\beta_{nl}$ are given by:
\begin{align}
\beta_{n0}&=2 \left[\ln(n)+\frac{1}{n}-\psi(n) \right], \\
\beta_{n1}&= \alpha_{n1} \beta_{n0}-\frac{4}{n^3},
\end{align}
where $\psi$ is the digamma function \cite{abramowitz+stegun}. Eq. \eqref{Tr} only involves the scattering length while third-order corrections require to compute the effective range $r_l$ and include terms proportional to $\left(\frac{a_l}{B^{2l+1}}\right)^3$ as well as mixed terms involving both $\left(\frac{a_l}{B^{2l+1}}\right)^2$ and $\left(\frac{r_l}{B^{2l+1}}\right)$. Their expressions are provided explicitly in Ref. \cite{T61}. It is worth noting that, for the $1s$ state, the first-order Trueman formula is analogous to the standard Deser formula \eqref{D0}, but involving the Coulomb-corrected scattering length $a_0$ instead of $A_0$. The series provided by higher-order terms will only converges when $a_l \ll  B^{2l+1}$, which typically holds for physical systems of interest, such as exotic hydrogen or deuterium.

\section{Three-body formalism}

\subsection{Faddeev equations}
Kaonic and antiprotonic deuterium atoms are modeled as three-particle systems interacting through pairwise interactions. The wavefunction includes two particle channels:  ($p$, $n$, $x$) and ($n$, $n$, $y$), which are coupled together by a charge-exchange process. Our objective is to compute the solution to the three-body Schr\"odinger equation:
\begin{equation}
\left(E-H_0-V_{NN}-V_{NX}-V_{nX} \right) \ket{\Psi} = 0 \quad \text{ with } \quad
 \ket{\Psi} \equiv \begin{pmatrix}
 \ket{\Psi_{pnx}} \\ \ket{\Psi_{nny}}
 \end{pmatrix},
\end{equation}
where $E$ is the energy, $H_0$ is the three-body kinetic energy operator, and the two-body potentials read
\begin{equation}
V_{NN} = \begin{pmatrix}
V_{pn,pn} & 0 \\
0 & V_{nn,nn}
\end{pmatrix}, \quad V_{NX} = \begin{pmatrix}
V_{px,px} & V_{px,ny} \\
V_{px,ny} & V_{ny,ny}
\end{pmatrix}, \quad V_{nX} = \begin{pmatrix}
V_{nx,nx} & 0 \\
0 & V_{ny,ny}
\end{pmatrix}.
\end{equation}
The $NX$ and $nX$ interactions are modeled using optical potentials to account for the annihilation of the particle $x$ with a nucleon. While typically defined in isospin space, their expression in the particle basis reads:
\begin{align}
V_{px} &= \frac{1}{2} \left(V^{(T=0)}_{NX}+ V^{(T=1)}_{NX} \right) + V_{C_{px}}, \\
V_{ny} &= \frac{1}{2} \left(V^{(T=0)}_{NX}+ V^{(T=1)}_{NX} \right), \\
V_{px \to ny} &= \frac{1}{2} \left(V^{(T=0)}_{NX}- V^{(T=1)}_{NX} \right), \\
V_{nx}&= V^{(T=1)}_{XN}.
\end{align}
Here $V_{C_{px}}$ is the $px$ attractive Coulomb potential. 

\begin{figure}[h]
    \centering
    \includegraphics[width=1.0\textwidth]{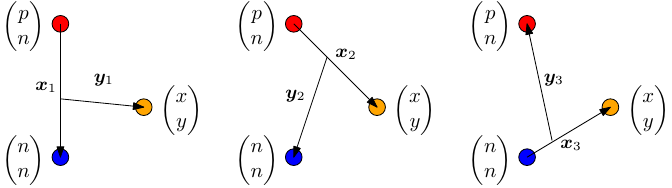}
    \caption{Jacobi coordinates for the $xd$ system.}
    \label{fig:jac123v}
\end{figure}

The three-body wavefunction is computed by solving the Faddeev-Merkuriev equations in configuration space, following the same formalism as described in Refs. \cite{CL21,DLD23}. The wavefunction is decomposed into three components as follows:
\begin{equation}
\Psi = \begin{pmatrix}
 \psi_{pn,x} \\ \psi_{nn,y}
 \end{pmatrix}(\bm{x}_1,\bm{y}_1)   + \begin{pmatrix}
 \psi_{px,n} \\ \psi_{ny,n}
 \end{pmatrix}(\bm{x}_2,\bm{y}_2)   +\begin{pmatrix}
 \psi_{nx,p} \\ \psi_{ny,n}
 \end{pmatrix}(\bm{x}_3,\bm{y}_3),
\end{equation}
where $\bm{x}_i$ and $\bm{y}_i$ are the Jacobi coordinates for each $2+1$ partition (see Figure \ref{fig:jac123v}) defined with the convention of Ref. \cite{DLD23}. Both bound and scattering states can then be computed by solving the following Faddeev-Merkuriev equations \cite{F60,M80}:
\begin{align}
\left(E-H^{ijk}_0- \delta m_{ijk} c^{2} - V^{(l)}_{C_{ik}} - V^{(l)}_{C_{jk}} \right) \psi_{ij,k} &= \sum_{i'j'} V_{ij,i'j'} \left(\psi_{i'j',k}+\psi_{i'k,j'} + \psi_{j'k,i'} \right) \nonumber \\
& \quad -V^{(l)}_{C_{ij}} \left(\psi_{ik,l} + \psi_{jk,i} \right), \label{Fad}
\end{align}
where $(i,j,k)$ represent the particles of a given channel, $H^{ijk}_0$ is the associated three-body kinetic energy operator, and the term $\delta m_{ijk} c^2$ accounts for the mass difference between the particle channels. The term $V^{(l)}_{C_{ij}}$ represents the long-range contribution of the Coulomb potential between particles $i$ and $j$, and is defined as
\begin{equation}
V^{(l)}_{C_{ij}}(x_{ij},y_{ij}) = V_{C_{ij}}(x_{ij}) \, \left[1-\chi(x_{ij},y_{ij}) \right], 
\end{equation}
with $\chi$ some cut-off function introduced to ensure the asymptotic decoupling of Faddeev-Merkuriev components ($\psi_{ij,k}$). 

\subsection{Numerical resolution}
The three-body states are characterized by their total angular momentum $J$ and  parity $\Pi$. To solve the Faddeev equations, each Faddeev component is represented through a partial wave decomposition, given by
\begin{equation}
\psi_{\alpha}(\bm{x}_{\alpha},\bm{y}_{\alpha}) = \sum_{n} \frac{\phi_{\alpha n}(x_{\alpha},y_{\alpha})}{x_{\alpha} y_{\alpha}} \, \mathcal{Y}_{n}(\hat{x}_{\alpha},\hat{y}_{\alpha}),  \label{PWE}
\end{equation}
where $n=\{l_x,s_x,j_x,l_y,j_y\}$ is a global index that encompasses a set of quantum numbers, $\phi_{\alpha n}$ is the radial wavefunction associated with the $n$-th amplitude, and $\mathcal{Y}_{n}$ represents the bispherical harmonics, which includes spin coupling. By inserting Eq. \eqref{PWE} into the Faddeev-Merkuriev equations and projecting onto a state of the angular momentum basis $\mathcal{Y}_n$, a set of coupled differential equations is obtained. The solution to these equations is carried out by expanding the radial wavefunctions over a two-dimensional Lagrange mesh \cite{B15}. 

The calculation of level shifts from the effective range parameters requires to solve the $d+x$ scattering problem at low energy. This is accomplished by applying the appropriate boundary conditions to the component $\psi_{pn,x}$ and solving the resulting non-homogeneous problem. Aside from their non-square-integrable asymptotic parts, the radial wavefunctions are expressed as a linear combination of products of Lagrange-Laguerre functions:
\begin{equation}
\phi_{n \alpha}(x_{\alpha},y_{\alpha}) = \sum_{i_x=1}^{k_x} \sum_{i_y=1}^{k_y} c_{n \alpha i_x i_y} \, f_{i_x}(x_{\alpha}) \, f_{i_y}(y_{\alpha}), \label{phi_coll}
\end{equation}
where $f_i(r)$ is a Lagrange-Laguerre function regularized by $\sqrt{r}$, defined over the interval $r \in [0,\infty[$. The basis is chosen to effectively capture the short-range interaction effects, with the grid extent primarily determined by the range of the nuclear interaction.

Alternatively, the level shifts can be computed directly by calculating 
$dx$ Rydberg states using bound-state techniques. However, due to the large spatial extension of these states, a standard Lagrange-Laguerre mesh struggles to accurately describe both the short- and long-range features of the wavefunction. This issue is addressed by dividing the configuration space into two domains along the $y$ coordinate, so the radial wavefunctions are defined as:
\begin{align}
\phi_{\alpha n}(x_{\alpha},y_{\alpha}) = 
\left\{
    \begin {aligned}
         &   \sum_{i_x=1}^{k_x} \sum_{i_y=1}^{\widetilde{k}_y} \, \widetilde{c}_{\alpha n i_x i_y} \, f_{i_x}(x_{\alpha}) \, \widetilde{f}_{i_y}(y_{\alpha})   \quad & (y_{\alpha} \leq R), \\
         & \sum_{i_x=1}^{k_x} \sum_{i_y=1}^{\hat{k}_y} \, \hat{c}_{\alpha n i_x i_y} \, f_{i_x}(x_{\alpha}) \, \hat{f}_{i_y}(y_{\alpha})  \quad & (y_{\alpha} > R).                  
    \end{aligned}
\right.
\end{align}
where $\widetilde{f}_i(r)$ is a Lagrange-Jacobi function defined over the interval $r \in [0,R]$ and $\hat{f}_i(r)$ is a shifted Lagrange-Laguerre function defined over the interval $r \in [R,\infty[$. This approach allows for an accurate numerical computation of the eigenvalues of the three-body Hamiltonian, providing numerically exact level shifts. However, it is heavier than scattering calculations from a numerical standpoint.

\subsection{Level shifts and quadrupole interaction}
When performing three-body calculations, the energies of the Coulomb orbits are computed relative to the deuteron energy $E_d \approx -2.22 \, \rm{MeV}$ and are expressed as
\begin{equation}
E^{(C)}_{n} = E_d + \epsilon_n, \label{Ecd}
\end{equation}
where $\epsilon_n$ is given by Eq. \eqref{eps_n} with $\mu$ the reduced mass of the $d+x$ system. The level shifts are defined as:
\begin{equation}
\Delta E_{nl}=E_{nl}-E^{(C)}_n,
\end{equation}
with $E_{nl}$ being the energy of the three-body state. Eq. \eqref{Ecd} assumes a point-like  Coulomb interaction between the particle $x$ and the deuteron center-of-mass. However, the actual $xd$ long-range interaction depends on the  deuteron charge distribution  and can be expressed as a multipole expansion:
\begin{align}
V_{C_{xd}} = -\frac{e^2}{r_{p x}} 
= -\frac{e^2}{s y_1} \sum_{l=0}^{\infty} \frac{4 \pi}{2l+1} \left(\frac{c x_1}{s y_1} \right)^l \bm{Y}^{(l)}(\hat{x}_1) \cdot \bm{Y}^{(l)}(\hat{y}_1), \quad (|s\bm{y}_1| > |c\bm{x}_1|). \label{pole}
\end{align}
By neglecting $l>2$ terms in this expansion, the asymptotic $xd$ interaction includes, in addition to the dominant Coulomb term (which scales as $\frac{1}{y_1}$ for $l=0$), a quadrupole term ($l=2$) that decreases as $\frac{1}{y^3_1}$. Therefore, the effective EM interaction becomes:
\begin{equation}
V_{xd}(y_1)  \xrightarrow[y_1 \to \infty]{} -\frac{e^2}{s y_1} + V_Q. \label{asymp}
\end{equation}
The potential $V_Q$ is generated by the $SD$ coupling in the deuteron wavefunction, which results in a non-spherical charge distribution and a non-zero quadrupole moment ($Q_d \approx 0.286 \, \rm{fm}^2$). These effects are dynamically included in bound-state calculations, but the resolution of scattering problems is more challenging due to the long-range coupling between different partial waves induced by terms of the form $\frac{1}{y^3_1}$. 

In systems described with central $NN$ interactions, $Q_d=0$ and quadrupole terms can be ignored for all states. However, for realistic $NN$ potentials, the contribution of $V_Q$ is zero only for three-body  $S$ states, while noticeable effects may appear for $P$ waves. In this case, the scattering length is undefined, and Eq. \eqref{effrange} is no longer valid, meaning the Trueman relation cannot be directly applied to compute the level shifts.

\subsection{Treatment of $P$ waves} \label{Pwaves}
In this section, we propose a method to approximately evaluate the $P$-wave level shifts from scattering observables, while maintaining the same formalism used for $S$ waves. The key idea is that the shift induced by the quadrupole interaction is several orders of magnitude smaller than the energy of the hydrogenic levels, which means that these effects can be treated perturbatively \cite{P11}. As a results, the level shift can approximately be written as
\begin{equation}
\Delta E_{nl} = \Delta E_N + \Delta E_Q,
\end{equation}
where $\Delta E_N$ and $\Delta E_Q$ are respectively the shifts induced by the nuclear and quadrupole potentials. The coupling of the deuteron spin and the spin of particle $x$ defines the hyperfine structure of the hydrogenic levels. Let us consider a spectrum including $g$ hyperfine levels $\ket{\Psi_i}$ ($i=1,...,g$) with total angular momentum $J$. Within the present perturbative approach, the unperturbed states are the Rydberg states only shifted by the strong interaction $\ket{\Phi_i}$ with energy $E^{(C)}_{n}+\Delta E_{N_i}$. These levels are computed similarly to $S$ states by solving a low-energy scattering problem but in which the quadrupole potential projected onto the deuteron wavefunction $\phi_d$ has been retrived from the three-body Hamiltonian. The corresponding Faddeev-Merkuriev equation for the component $\psi_{pn,x}$ then reads
\begin{align}
\left(E-\Delta E_Q-H_0-V_{pn}-V^{(l)}_{C_{px}} \right) \ket{\psi_{pn,x}} &= V_{pn} \left( \ket*{\psi_{p x,n}} + \ket*{\psi_{nx,p}} \right) - W \ket*{\Psi_{pnx}}, \label{Fadmod}
\end{align}
where the operator $W$ is defined as
\begin{equation}
    W = \ket*{\phi_d} \mel*{\phi_d}{V_Q}{\phi_d} \bra{\phi_d}.
\end{equation}
The energy shifts due to the nuclear interaction are then estimated from the calculated scattering length. For numerical convenience, the quadrupole potential is regularized near the origin by the function $(1-e^{-y/y_c})^3$ with $y_c \approx 1 \, \rm{fm}$.

The following step consists in calculating the shift induced by the quadrupole potential. By considering 
\begin{equation}
\ket{\Psi_i} = \sum_{i=1}^g c_i \ket{\Phi_i}, 
\end{equation}
the quadrupole corrections are evaluated at the first order of perturbation theory by solving the eigenvalue problem given by
\begin{equation}
\sum_{j=1}^g \mel{\Phi_i}{V_Q}{\Phi_j} c_i = (E-E^{(C)}_n-\Delta E_{N_i}) c_i. \label{EG}
\end{equation}
Outside of the range of the nuclear interaction, only electromagnetic effects play a significant role. For the calculation of the left-hand side of Eq. \eqref{EG}, the three-body wavefunction is approximated by the product of the deuteron wavefunction and the hydrogenic wavefunction $\Psi^{(0)}_{nl}$ corresponding to the energy $\epsilon_n$:
\begin{equation}
\ket{\Phi_i} \approx \ket{\phi_d \Psi^{(0)}_{nl}}.
\end{equation}
 These functions are well-known and the left-hand side of Eq. \eqref{EG} can be calculated analytically using \cite{BS_book}:
\begin{equation}
\int_{0}^{\infty} \frac{|\Psi^{(0)}_{nl}(y)|^2}{y} \, \mathrm{d} y = \frac{1}{l(l+\frac{1}{2})(l+1)n^3}.
\label{qpef}
\end{equation}
Worth noting that energy shifts due to short-range interactions decrease exponentially with $l$, while quadrupole moment corrections scale only as  $1/l^3$, the quadrupole correction becomes dominant for high angular momentum values.

\section{Results}

\subsection{A two-body example}
As an initial application, we examine the accuracy of the Deser and Trueman formulas in a simple two-body system, similar to the model used in Ref. \cite{MHR20}. The particles are supposed to interact through both short- and long-range potentials, defined as follows:
\begin{align}
V(r) &= \lambda \, V_{\rm{sr}}(r) + V_{C}(r), \\
V_{\rm{sr}}(r) &= - C \, e^{- \nu r^2}, \\
V_{C}(r) &= - \frac{1}{r}, \label{V2}
\end{align}
with parameters $C=2420$ and $\nu=30$. The particle masses and physical constants are set to $1$, and $C$ is chosen such that the short-range potential alone supports a bound state for  $\lambda>~1$. The Rydberg states are characterized by $\epsilon_n=-\frac{1}{4n^2}$ and $B=2$. 

\begin{figure}[h]
    \centering
    \includegraphics[width=1.0\textwidth]{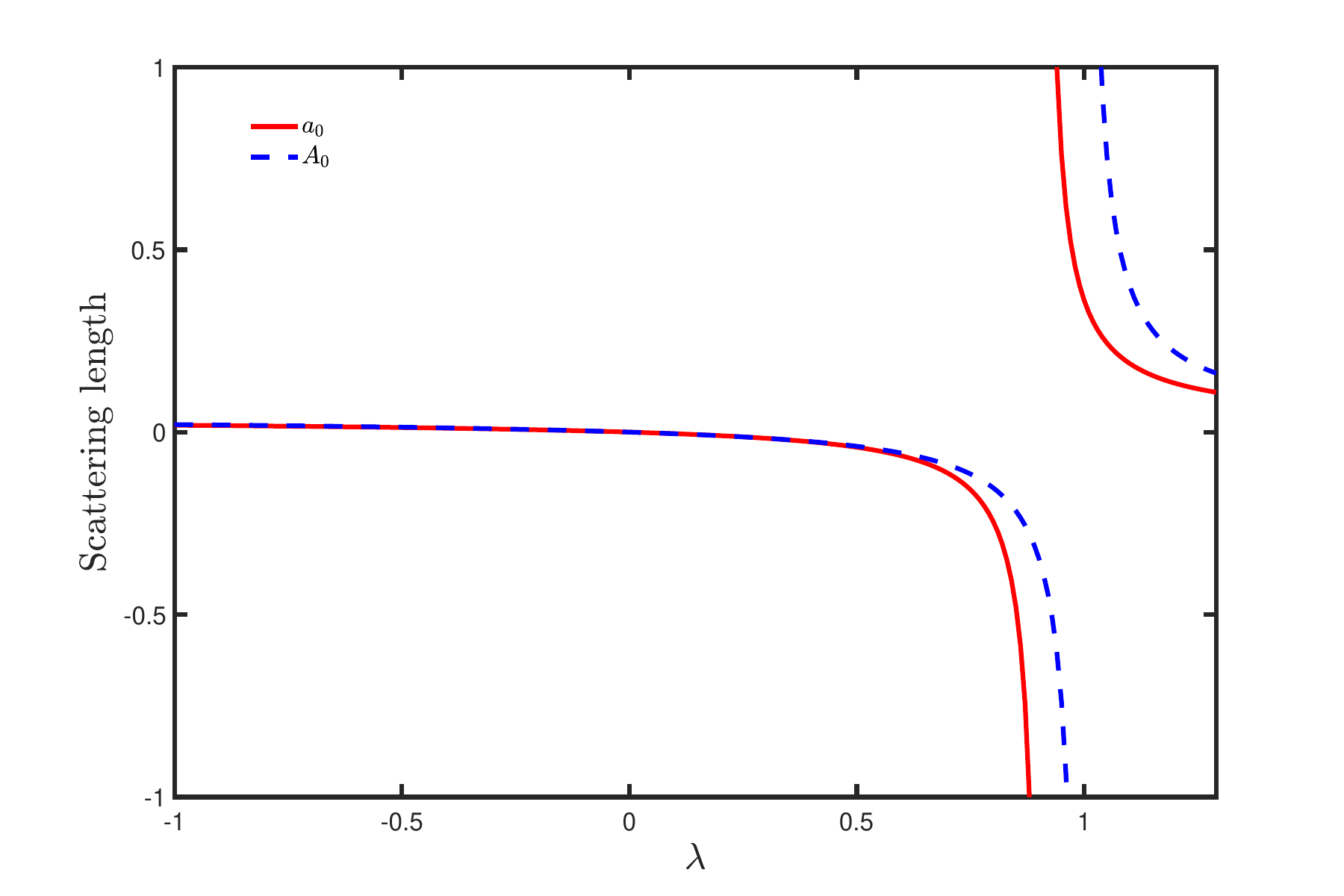}
    \caption{$S$-wave scattering length calculated with (solid line) and without (dashed line) accounting for the Coulomb potential, as a function of $\lambda$.}
    \label{fig:Aa0}
\end{figure}

Figure \ref{fig:Aa0} shows the $S$-wave scattering lengths calculated with and without the Coulomb potential, as a function of $\lambda$. For the short-range interaction alone, the scattering length is positive for $\lambda<0$ and negative for $\lambda>0$. As $\lambda$ approaches $1$, $A_0$ becomes infinite due to the emergence of a bound state.  When considering the full potential in Eq. \eqref{V2}, the scattering length $a_0$ follows a similar trend to $A_0$ until around $\lambda \approx 0.5$, after which it approaches $-\infty$ around $\lambda \approx 0.9$ due to the additional attractive force provided by the Coulomb potential. 

\begin{figure}[h]
    \centering
    \includegraphics[width=1.0\textwidth]{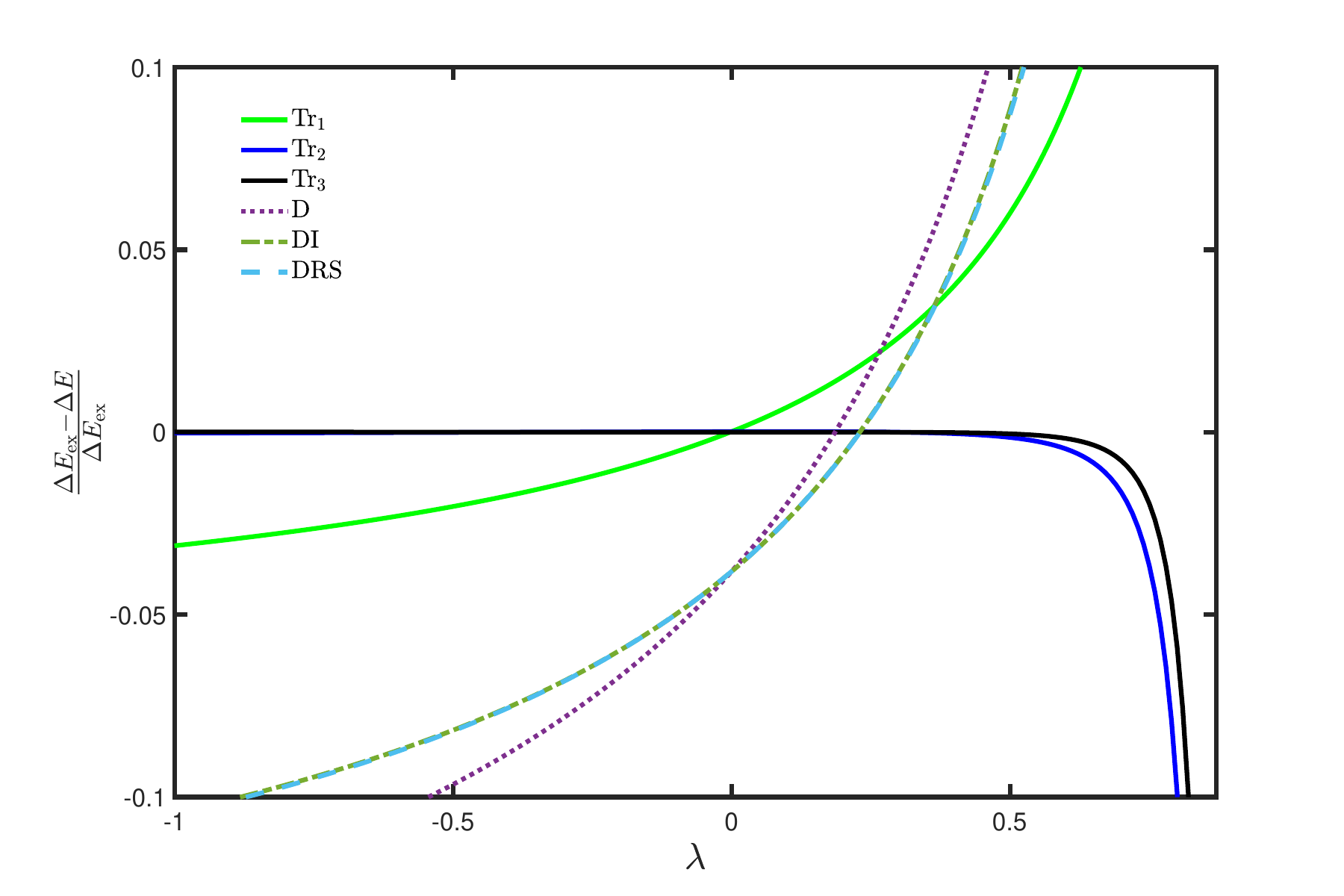}
    \caption{Relative difference between the exact $1s$ level shift, and the one computed with the standard (D), improved (DI), and resummed (DRS) Deser formulas, and with the Trueman relation from first to third order ($\rm{Tr}_1$, $\rm{Tr}_2$, and $\rm{Tr}_3$ respectively). The DI and DRS curves are almost superimposed.}
    \label{fig:Derel}
\end{figure}

The $1s$ level shift has been calculated using both the Deser formulas \eqref{D0}-\eqref{DRSe}, and from the Trueman relation \eqref{Tr} up to third order. Figure \ref{fig:Derel} presents the relative difference between the approximated and exact level shifts as a function of $\lambda$. 

Compared to the standard Deser formula \eqref{D0}, the improved Deser formula provides more accurate results, except in the region with $0 \leq \lambda \leq 0.2$, where the curves change sign. The resummed Deser formula does not significantly enhance the accuracy, yielding results almost indistinguishable from the improved Deser formula. For most values of $\lambda$, the relative difference from the exact level shift remains within a few percent. Interestingly, the Deser formulas do not necessarily yield better relative accuracy near  $\lambda=0$, where the short-range potential has the weakest influence on the Coulomb-bound states. 

Trueman relation up to first order is accurate up to $1\%$ in a narrow region near $\lambda=0$ from where it starts deviating. However, the inclusion of second- and third-order terms improves systematically accuracy across all values of $\lambda$. As $\lambda \to 0.9$, the scattering length increases, slowing the convergence of the series. For $\lambda>0.85$, one has $a_0>B$ and the series diverges, leading to inaccurate results. Outside of this region, the inclusion of higher-order terms consistently yields relative differences below $1\%$ compared to the exact values.  When applicable, the Trueman relation—especially with higher-order terms—should be preferred over the Deser formulas.

An alternative to using the Trueman relation is to numerically solve the non-linear equation provided by  Eq. \eqref{effrange} truncated at the quadratic term. This approach produces, for any value of $\lambda$, highly accurate results indistinguishable from those obtained through direct bound-state energy calculations, within the limits of our attained numerical precision.

\subsection{Kaonic deuterium ($K^{-}d$)}
We now consider the case of kaonic deuterium, in which the electron of the deuterium atom is replaced by an antikaon $K^-$. The three-body problem explicitly includes the particles channels ($p$,$n$,$K^-$) and ($n$,$n$,$K^0$) coupled together by a charge exchange process. This system has for instance been investigated  in Refs. \cite{HOHHW17,Revai16}. The nucleon-antikaon ($N \bar{K}$) interaction is modeled with the optical, energy-dependent Kyoto potential \cite{KyotoPot}, whose parameters have been adjusted to reproduce the level shifts observed in kaonic hydrogen, as measured by the SIDDHARTA experiment \cite{SIDD}. It was demonstrated in Ref. \cite{HOHHW17} that the effects of the energy-dependence can be neglected, and hence $E_{N\bar{K}}=0$ is used in practice.

We compare different $NN$ interactions: the phenomenological Malfliet-Tjon (MT-I-III) \cite{MT13} and Minnesota (exchange parameter $u=2$) \cite{Minnesota} potentials, which are both central, as well as the realistic AV18 potential \cite{AV18}. Similarly to Ref. \cite{HOHHW17}, the calculations have been carried out considering the physical masses $m_p=938.272 \, \frac{\rm{MeV}}{c^2}$, $m_n=939.565 \, \frac{\rm{MeV}}{c^2}$, $m_{K^-}=493.677 \, \frac{\rm{MeV}}{c^2}$, $m_{K^0}=497.648 \, \frac{\rm{MeV}}{c^2}$, and the following constants: $\hbar c = 197.32698 \, \rm{MeV \, fm}$ and $\alpha = 1/137.0360$. All partial waves with $l_x,l_y \leq 6$ have been included, which was found sufficient for the convergence. 

Table \ref{tab:kdAa} summarizes the effective range parameters of the $K^-d$ system   for the  $J^{\Pi}=1^+$ state. From these values, the $1s$ level shift is computed using both the Deser and Trueman formulas, and compared to the result obtained from a bound-state approach.

\begin{table}[h]
\caption{\label{tab:kdAa}%
$K^-d$ effective range parameters computed with different $NN$ interactions in conjunction with the Kyoto potential.
}
\begin{ruledtabular}
\begin{tabular}{cccc}
 & \multicolumn{1}{c}{MT-I-III}  & \multicolumn{1}{c}{Minnesota} & \multicolumn{1}{c}{AV18} \\
\colrule
$A_0 \, \rm{(fm)}$  & $1.404-1.35 \, $i & $1.296-1.251 \, $i & $1.45-1.361 \, $i \\
$a_0  \, \rm{(fm)}$ & $1.303-1.040 \, $i & $1.207-0.979 \, $i & $1.34-1.044 \, $i \\
 $r_0 \, (\rm{fm})$ & $0.778-1.37 \, $i & $0.720-1.419 \, $i & $0.793-1.27 \,$i \\
\end{tabular}
\end{ruledtabular}
\end{table}

Table \ref{tab:kdDE} demonstrates that the Deser formulas are unable to reproduce the exact results. Notably, the values from both the standard and improved Deser formulas differ significantly from those obtained through direct calculations. The better results are obtained using the resummed Deser formula, which yields only a few percents difference from the exact values.
Due to discrepancy between $A_0$ and $a_0$ -- especially in their imaginary parts -- which arises from the inclusion of the Coulomb potential, the first-order Trueman formula exhibits less error than both the standard and improved Deser formulas. At second order, the Trueman relation achieves excellent agreement with the exact result, with a deviation  of approximately $1 \, \text{eV}$. The third-order correction reduces the error only slightly but involves the additional effort of computing the effective range besides the scattering length.

We also note discrepancies between our results and those reported in Ref. \cite{HOHHW17}. Using the Minnesota potential for the $NN$ interaction, the authors of this reference calculate the $1s$ level shift employing a basis of correlated Gaussian functions, getting $\Delta E_{1s}=(670-508 \, i) \, \rm{eV}$, which is slightly smaller than the value listed in Table \ref{tab:kdDE}. The observed differences may be attributed to distinct versions of Minessota and $N \bar{K}$ potentials employed in our work, as the authors of Ref. \cite{HOHHW17} were enable to provide details of their implementation. Furthermore, they extracted the value of the scattering length $A_0=(1.42-1.60 \, i) \, \rm{fm}$ using a multiple-scattering series approximation, which only relies on  the two-body scattering lengths and the deuteron density. This approximation  is based on a strongly simplified picture of three-body dynamics and leads to a 20\% deviation from the result of the full three-body calculation. It further affects the accuracy of the Deser result reported in Ref. \cite{HOHHW17}. 

While the Deser formulas are limited to the calculation of the $1s$ level shift, the Trueman relation provides a systematic method to determine the level shifts of both ground and excited states using the same set of parameters. For example, the $2s$ level shift, computed using Eq. \eqref{Tr} up to third order, is presented in Table \ref{tab:kdDE2s} and compared to the value obtained from a direct calculation. Unlike the $1s$ case, the first-order Trueman formula already provides a reasonable approximation for the  $2s$ level shift, particularly for its real part. The result further improves by second and third-order corrections and is less than $1 \, \text{eV}$ off from the exact value.

Finally, regarding the dependence on the $NN$ interaction, the results obtained using the MT-I-III and AV18 potentials are quite similar, despite their different descriptions of the deuteron wavefunction. For the three $NN$ potentials considered, the maximum relative differences are about 10\% for the real part and 5\% for the imaginary part of the level shifts.


\begin{table}[h]
\caption{\label{tab:kdDE}%
The $1s$ $K^-d$ level shifts computed from the standard ($\Delta E^{(D_0)}_{1s}$), improved ($\Delta E^{(D_I)}_{1s}$), and resummed Deser ($\Delta E^{(D_{\Sigma})}_{1s}$) formulas, from Trueman relation at first ($\Delta E^{(T_1)}_{1s}$), second ($\Delta E^{(T_2)}_{1s}$), and third ($\Delta E^{(T_3)}_{1s}$) order, and from direct bound-state calculations ($\Delta E_{1s}$). 
}
\begin{ruledtabular}
\begin{tabular}{cccc}
$1s \, (J^{\Pi}=1^+)$ & \multicolumn{1}{c}{MT-I-III}  & \multicolumn{1}{c}{Minnesota} & \multicolumn{1}{c}{AV18} \\
\colrule
Deser formulas & & & \\
$\Delta E^{(D_{0})}_{1s} \, (\rm{eV})$  & $845-813 \, $i & $780-753 \, $i & $873-819 \, $i \\
$\Delta E^{(D_{R})}_{1s} \, (\rm{eV})$   & $830-422 \, $i & $768-419 \, $i & $847-413 \, $i \\
$\Delta E^{(D_{\Sigma})}_{1s} \, (\rm{eV})$ & $776-510 \, $i & $724-489 \, $i & $794-508 \, $i \\  
\colrule
Trueman formulas & & & \\
$\Delta E^{(T_1)}_{1s} \, (\rm{eV})$   & $784-626 \, $i & $726-589 \, $i & $806-628 \, $i \\
$\Delta E^{(T_2)}_{1s} \, (\rm{eV})$   & $767-552 \, $i & $713-524 \, $i & $787-552 \, $i \\
$\Delta E^{(T_3)}_{1s} \, (\rm{eV})$  & $768-549 \, $i & $713-522 \, $i & $788-549 \, $i \\  
\colrule
$\Delta E_{1s} \, (\rm{eV})$ & $768-550 \, $i & $713-523 \, $i & $787-550 \, $i \\  
\end{tabular}
\end{ruledtabular}
\end{table}

\begin{table}[h]
\caption{\label{tab:kdDE2s}%
The $2s$ $K^-d$ level shifts computed from the Trueman formula at first ($\Delta E^{(T_1)}_{2s}$), second ($\Delta E^{(T_2)}_{2s}$), and third ($\Delta E^{(T_3)}_{2s}$) order, compared with the values obtained from direct bound-state calculations ($\Delta E_{2s}$). 
}
\begin{ruledtabular}
\begin{tabular}{cccc}
$2s \, (J^{\Pi}=1^+)$  & \multicolumn{1}{c}{MT-I-III}  & \multicolumn{1}{c}{Minnesota} & \multicolumn{1}{c}{AV18} \\
\colrule
$\Delta E^{(T_1)}_{2s} \, (\rm{eV})$   & $98.0-78.2 \, $i & $90.8-73.6 \, $i & $100.8-78.5 \, $i \\
$\Delta E^{(T_2)}_{2s} \, (\rm{eV})$   & $97.0-73.7 \, $i & $89.9-69.7 \, $i & $99.6-73.9 \, $i \\
$\Delta E^{(T_3)}_{2s} \, (\rm{eV})$ & $97.4-72.9 \,$i & $90.3-69.0 \,$i & $100-73.0 \,$i \\
\colrule
$\Delta E_{2s} \, (\rm{eV})$ & $97.3-73.1 \, $i & $90.2-69.2 \, $i & $99.9-73.2 \, $i \\  
\end{tabular}
\end{ruledtabular}
\end{table}

\subsection{Antiprotonic deuterium ($\bar{p}d$)}
Within the optical model framework, the $\bar{p}d$ system involves the following particles channels: ($p$,$n$,$\bar{p}$) and ($n$,$n$,$\bar{n}$). The $N \bar{N}$ interaction is modeled using the Kohno-Weise optical potential \cite{KW86}, while both central and realistic $NN$ interactions are considered. This system was previously studied in Ref. \cite{CL21} where the level shifts of $S$ and $P$ states were computed using  bound-state technique. While the models agreed well for $S$ waves, strong discrepancies were found for coupled $P$ 
waves. Specifically, when the Malfliet-Tjon potential was used in conjunction with the Kohno-Weise potential, the level shifts for the $P^{-}_{1/2}$ and $P^{-}_{3/2}$ states were notably different from those obtained with realistic $NN$ interactions.

The Trueman relation was previously applied only to uncoupled waves, where the effects of quadrupole interactions were either negligible or absent. In the present work, we extend this approach to coupled $P$-waves by applying the methodology detailed in section \ref{Pwaves}. The three-body calculations are performed using the nucleon mass $m_N=938.94 \, \frac{\rm{MeV}}{c^2}$ and including all partial waves with $l_x,l_y \leq 6$.

Table \ref{table:pwave_pbard}  presents the level shifts of various
 $P$ states computed from a direct bound-state approach and from the second-order Trueman relation, for different $NN$ potentials. The hyperfine structure of the $\bar{p}d$ system includes two $P$ states with $J^{\pi}=\frac{1}{2}^-, \frac{3}{2}^-$ and one with $J^{\Pi}=\frac{5}{2}^-$. In all cases, the results obtained from the first- and second-order Trueman relations are indistinguishable within the limits of our numerical precision, allowing us to neglect third-order corrections. 
 
For the MT-I-III potential, which does not include $SD$-coupling in the deuteron wavefunction, the level shift arises solely from the nuclear interaction ($Q_d=0$, $V_Q=0$). In contrast, calculations using the AV18 potential account for both nuclear and quadrupole contributions ($Q_d \approx 0.270 \, \text{fm}^2$, $V_Q \neq 0$). The quantity $\Delta E_{np}$ represents the level shift computed from the bound state approach. The quantity $\Delta E^{(T_2)}_{np}$ represents the hadronic shift computed by solving Eq. \eqref{Fadmod} and $\Delta E^{(T_2+V_Q)}_{np}$ is the full shift including the perturbative quadrupole effects. The quadrupole corrections lead to significant differences in the level shifts for the coupled $P_{1/2}$ and $P_{3/2}$ states.  A smaller, though still noticeable, effect is also observed in the uncoupled $P_{5/2}$ state. For both $NN$ potentials, the second-order Trueman relation provides a reliable estimate of the level shifts for the $2P$ and $3P$ states, based solely on the scattering length. The differences between the scattering and bound-state approaches can be attributed to two main factors: the accuracy of the bound-state approach limited to a few meV and the truncation of the electromagnetic interaction to the quadrupole term in scattering calculations.

When comparing the level shifts induced solely by nuclear interactions, the results obtained with the MT-I-III and AV18 potentials are quite similar. The largest differences appear for the states with $J^{\Pi}=\frac{1}{2}^-$ for which the quadrupole effects play a dominant role. Table \ref{table:pwave_pbard} also contains the spin-averaged level shifts. Despite of the fact that the absence of $SD$ coupling in the MT-I-III potential leads to very different results in comparison with realistic potentials, the spin-average value is rather $NN$ independent as the quadrupole corrections average to zero. As already pointed out in Ref. \cite{CL21}, strong discrepancies are observed between the predicted and the experimental data, especially concerning the real part of the energy shift. Moreover, a comparison with calculations performed with other $N \bar{N}$ interactions such as the Paris \cite{Paris09,LW20} or J\"ulich \cite{Julich,DLC23} potentials also suggests that the $P$-wave level shifts could be $N \bar{N}$ model-dependent.

\begin{table}[h]
\caption{\label{table:pwave_pbard}%
Level shifts of $d \bar{p}$ $P$ states computed from the direct calculation ($\Delta E_{np}$) and from the second order Trueman relation with ($\Delta E^{(T_2)}_{np}$) and without quadrupole corrections ($\Delta E^{(T_2+V_Q)}_{np}$), for different $NN$ interactions used in conjunction with the KW potential.
}
\begin{ruledtabular}
\begin{tabular}{c|cc|ccc}
$L^{\pi}_J$ & \multicolumn{2}{c|}{MT-I-III}  & \multicolumn{3}{c}{AV18} \\
   \colrule
&   $\Delta E^{(T_2)}_{np}$ (meV) & $\Delta E_{np}$ (meV) & $\Delta E^{(T_2)}_{np}$ (meV) & $\Delta E^{(T_2+V_Q)}_{np}$ (meV) & $\Delta E_{np}$ (meV)  \\
   \colrule
  $^2P_{1/2}$ ($n=2$) & $48.0-195\,$i & $41.5-195 \,$i & $13.8-188 \,$i & $-24.6-216 \,$i & $-31.4-215 \,$i\\
  $^4P_{1/2}$ ($n=2$) & $54.8-239 \,$i & $47.5-239 \,$i & $70.0-225 \,$i & $209-196 \,$i & $203-204 \,$i \\
  $^2P_{1/2}$ ($n=3$) & $16.7-68.4\,$i & $14.8-68.6 \,$i & $4.8-65.9 \,$i & $-3.1-75.8 \,$i & $-5.5-75.6 \,$i\\
  $^4P_{1/2}$ ($n=3$) & $19.3-83.8 \,$i & $16.4-84.0 \,$i & $24.6-79.0 \,$i & $62.2-69.0 \,$i & $57.8-72.2 \,$i \\
 \colrule
  $^2P_{3/2}$ ($n=2$) & $46.4-189\,$i & $39.7-189 \,$i & $43.0-187 \,$i & $61.6-189 \,$i & $53.0-190 \,$i\\
  $^4P_{3/2}$ ($n=2$) & $54.8-208 \,$i & $48.2-208 \,$i & $67.9-216 \,$i & $-30.9-214 \,$i & $-35.2-213 \,$i \\
  $^2P_{3/2}$ ($n=3$) & $16.3-66.4\,$i & $13.4-67.0  \,$i & $15.1-65.5 \,$i & $20.5-66.3 \,$i & $17.2-66.8\,$i\\
  $^4P_{3/2}$ ($n=3$) & $19.3-72.9 \,$i & $16.7-72.9  \,$i & $23.9-75.6 \,$i & $-5.3-75.0 \,$i & $-7.8-75.0 \,$i \\
  \colrule
   $^4P_{5/2}$ ($n=2$) & $51.6-208\,$i & $45.1-208 \,$i & $47.8-202 \,$i & $67.8-202 \,$i & $64.4-203 \,$i\\
   $^4P_{5/2}$ ($n=3$) & $18.1-73.1\,$i & $15.5-73.1 \,$i & $16.8-70.9 \,$i & $22.7-70.9 \,$i & $19.6-72.5 \,$i\\
   \colrule
   Spin-averaged ($n=2$) & $51.1-206 \,$i & $44.5-206 \,$i & $50.0-203 \,$i & $49.9-203 \,$i & $44.5-204 \,$i \\
   Spin-averaged ($n=3$) & $17.9-72.2 \,$i & $15.3-72.4 \,$i & $17.5-71.1 \,$i & $17.5-71.1 \,$i & $14.4-72.1 \,$i \\
\end{tabular}
\end{ruledtabular}
\end{table}

\section{Conclusion}

In this work, we reviewed several approaches for calculating the level shifts of exotic atoms by utilizing their connection to effective range parameters. Our primary focus was on evaluating the reliability and accuracy of the Deser and Trueman formulas, originally developed for two-body systems, and exploring their applicability to more complex, multi-body cases.
By solving the Faddeev-Merkuriev equations in configuration space, we computed the level shifts of kaonic and antiprotonic deuterium through both direct and indirect methods.

The Deser formulas directly connects the strong $S$-wave scattering length to the $1s$ level shifts.  However, because these formulas neglect interference effects between strong and electromagnetic interactions, they can lead to significant errors in predicting level shifts.  In contrast, Trueman’s expansion fully includes the Coulomb potential, making it a more reliable method. At second order, the Trueman formula requires only the scattering length and produces highly accurate results for both $S$- and $P$-waves. This indirect approach is computationally efficient, offering a promising alternative to traditional bound-state calculations, particularly for studying heavier exotic atoms.
  
  Nevertheless, if the nucleus possesses a quadrupole moment, special care should be taken to account for the EM interaction terms
  that decrease as $\frac{1}{y^3}$, which are not accounted for in the Trueman formula. For $P$  and higher partial waves, the  effect of $l=2$ and higher-order electromagnetic multipole terms on Rydberg level shifts may exceed the influence of hadronic interaction. Since this electromagnetic interaction is weak at short distances—where hadronic forces dominate—its contribution to the level shifts can be treated separately by subtracting the effective interaction terms from the total Hamiltonian. Effect of EM terms proportional to $\frac{1}{y^3}$  on energy shifts can be accurately evaluated by using first-order perturbation theory. This procedure has been successfully applied in calculating the energy shifts of antiproton-deuteron $P$ states.

\clearpage

\section{Acknowledgement}
This work has received funding from the F.R.S.- FNRS under Grant No. 4.45.10.08. One of the authors (P.-Y. Duerinck) is a Research Fellow at F.R.S.-FNRS.  We were granted access to the HPC resources of TGCC/IDRIS under the allocation 2024-AD010506006R3 made by GENCI (Grand Equipement National de Calcul Intensif). This work was supported by French IN2P3 for the PUMA project.


\providecommand{\noopsort}[1]{}\providecommand{\singleletter}[1]{#1}%

\end{document}